# Reactive nitrogen plasma spray coating of titanium nitride: plasma torch design and coating analysis

**Sina Mohsenian[1*], Jafar Fathi[2] and Babak Shokri[2]**

[1] Department of Mechanical Engineering, University of Massachusetts Lowell, Ma, USA.
Email: sina_mohsenian1@uml.edu
[2] Laser and Plasma Institute, Shahid Beheshti University, Tehran, Iran.

**Abstract.** In this paper, a thermal DC plasma torch was designed and developed for plasma spraying of a titanium nitride layer. Thermal plasma spray torch is a kind of non-transferred torch which can produce a high energy and high temperature plasma arc jet. Ti powders were injected into the plasma jet interacted with nitrogen plasma, resulted in a thick and hard layer of titanium nitride. Coating sample produced by this method was analysed with Raman spectroscopy confirming titanium nitride production. To study the thickness of the layer optical microscopy was used. It was found that the resulted layer is thicker than the layer synthesized by other deposition methods and it mainly consists of pure titanium nitride.

## 1. Introduction

Titanium Nitride (TiN) thin films because of high resistance against erosion, chemical corrosion, and high temperatures, and also low friction coefficient [1-3] are widely used in ceramic engineering especially thin films deposition on different cutting tools, semiconductors technology, microelectronic devices, solar cells, thermal barrier coatings, fuel container, combustion reactors and teeth orthopaedic proteases. Moreover, golden colour of titanium nitride makes it useful in protective and ornamental covers [4-21].

Nowadays, titanium nitride thin film deposition is mostly performed by physical vapor deposition (PVD) and chemical vapor deposition (CVD) methods [22-24]. However, the deposited layer in these methods are so thin, less than 10μm, and causes these technologies not to be considered industrially advantageous.

Thermal plasma torch is another method which can be applied for thin films deposition. Thermal plasma torch as an environmentally friendly device has many applications. It can provide an elevated temperature region with high fluxes of heat and reactive species [25-27]. This condition would be appropriate for melting micro-size metal particles, accelerating them as a stream of molten particles to substrate. Finally, liquid particles are flattened and solidified on the substrate, resulting in a formation of hard layers with thickness of a few ten micrometres [28].

Studies show that in industrial applications TiN deposited films with higher thickness is more desirable. TiN deposited films have better resistance against erosion and chemical corrosion. Furthermore, if TiN thin film thickness is less than 12μm, it is not able to resist against osmotic corrosion. Therefore, producing a thick TiN deposited film, considered a great technique for improving thin film efficiency. Applying plasma spray technology to produce a thick TiN film was reported in some publications. Bacci et al. [2] used inactive plasma spray deposition produced a 60μm TiN film sprayed in atmospheric nitrogen chamber at 500bar. However, his deposited thin film mainly was a combination of residues like Ti, TiN and Ti2N. Kobayashi et al. [1] used a gas tunnel plasma jet produced a 200µm TiN film. Also, Feng et al. studied micro resistance and tribology of a plasma sprayed TiN film [29].

In the present work, an atmospheric nitrogen DC thermal plasma torch has been designed and developed for making a TiN film on titanium surface. The torch operates in free air atmosphere with high electrical current (more than 100 A), in which the generated plasma is stabilized by swirl nitrogen flow. Section 2 describes the reactor design. Section 3 presents the experimental setup. The characteristics of deposited layer is presented in Section 5, including its thickness and composition.

## 2. Torch Design

The torch is designed to produce an atmospheric pressure high temperature plasma jet in local thermodynamic equilibrium mode. In this mode, the gas temperature of plasma jet is very high (more than 8000 K) which is appropriate for thermal spray.

The torch design, together with its main components and generated plasma jet, is shown in Figure 1. This torch is a kind of non-transferred torch which mainly consists of two co-axial water-cooled electrodes (anode and cathode). An AC high voltage power supply (30 kV, 30 MHz) was used as a trigger to strike the arc between electrodes. Then, a DC high current power supply (150 V-200 A) was applied to produce a high-power thermal plasma jet. A high-power water-cooled resistor (15 $\Omega$) was also used to stabilize the low intensity plasma arc which has a negative resistance characteristic. The arc voltage, arc current and subsequently power consumed in the plasma torch are 35 V, 140 A and 4.9 kW, respectively. After striking a DC high current arc between two electrodes, a vortex stream of nitrogen gas passes through this arc. This resulted in an elevated temperature and high velocity of thermal plasma jet. The gas flow rates were controlled by mass flow meter adjusted for nitrogen gas which is set on 5 slpm.

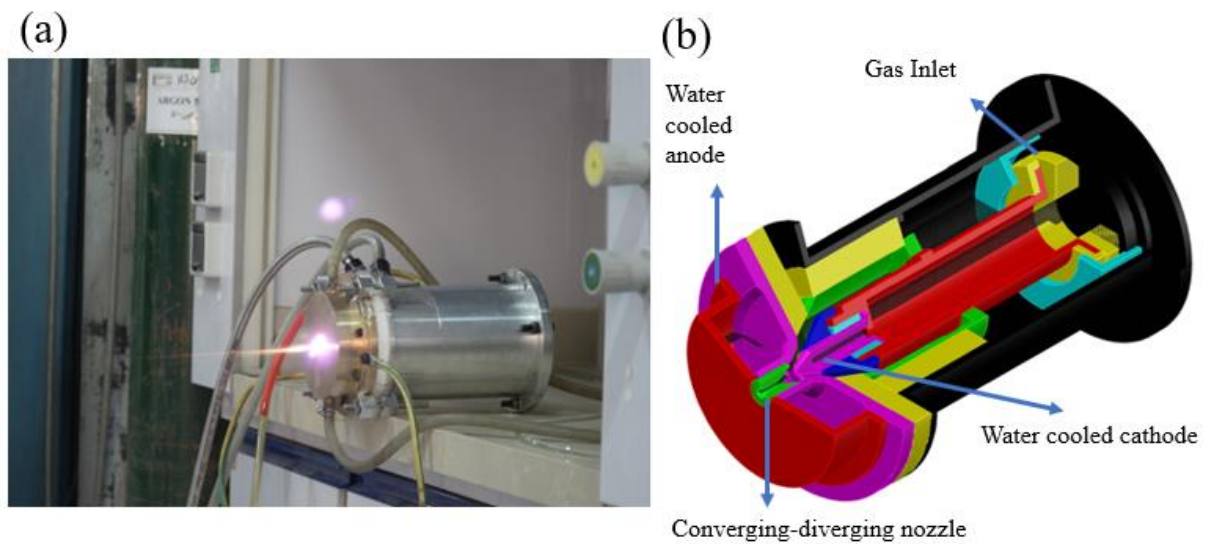

Figure 1: (a) picture of DC thermal plasma torch during operation, (b) illustration of components

A converging-diverging nozzle was used to channel the plasma to form a stable jet. The use of a conductive nozzle at the end of plasma torch prevents the divergence of the electric energy distribution, resulting in a stabilized and expanded plasma jet.

## 3. Experimental setup

The experimental setup of DC plasma spray torch and its auxiliary equipment used in this study are shown in Figure 2. This system comprises a DC thermal plasma torch, a high-power dc power supply, a high voltage ac ignitor, gas supplies and controller, a water cooling system, powder feeder as well as some diagnostic instruments.

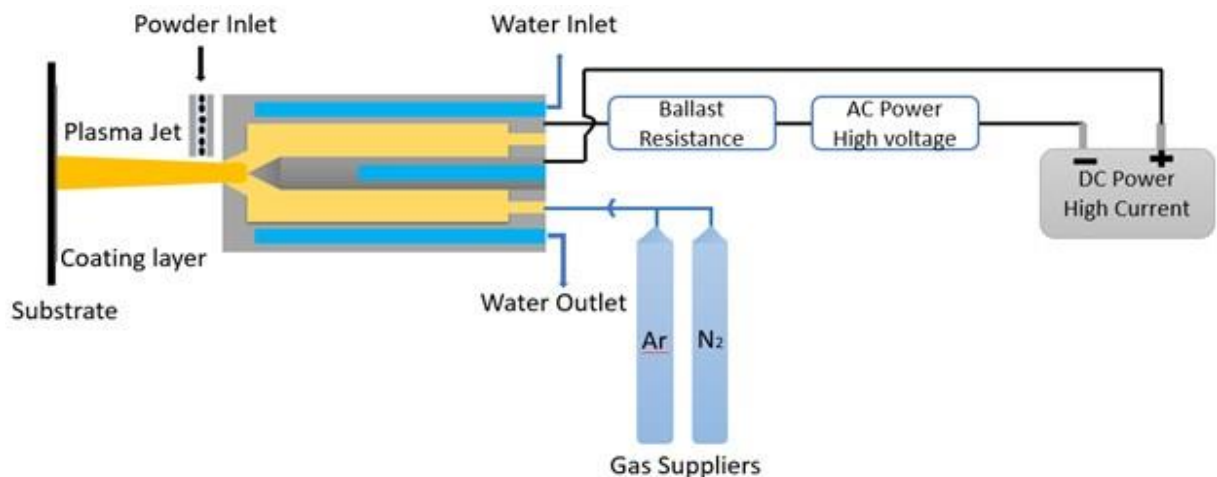

Figure 2: Schematic diagram of the experimental setup

Ti powders (grain size < 60 µm), were injected into a plasma jet with the rate of 1 mg/s. During the experiments, it has been found that this value for powder injection rate would be the best one for this torch to create a pure TiN layer on Ti substrate. Indeed, with this range, particles have enough time to be nitrated before reaching the substrate. Spraying distance was 100 mm.

A pure titanium plate with thickness of 2 mm, were used as substrate material. Due to elevated temperature of plasma jet, Ti particles react with nitrogen atoms before attaching to substrate to generate the TiN molecules. Consequently, the titanium nitride particles were attached on a titanium nitride substrate by high velocity of plasma jet. In this method, substrate was sprayed about 3 minutes to create a 40 µm layer. After process, the samples were polished to be prepared for analysis and measurements.

The resultant TiN layer were characterized using three techniques: optical microscopy (OM), Atomic force microscopy (AFM) and Raman spectroscopy (RS).

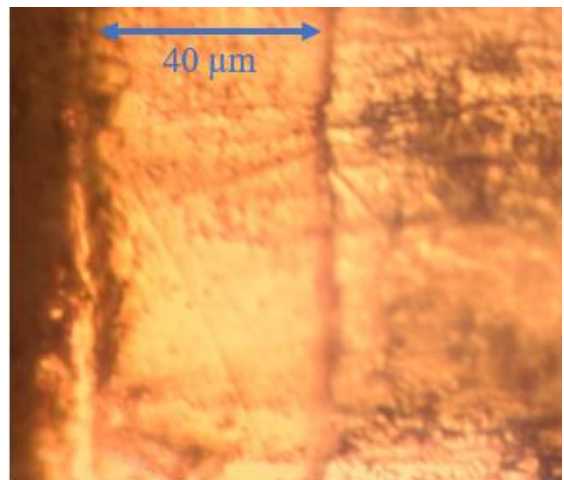

Figure 3: Cross surface of layer and substrate

## 4. Results and discussion

Figure 3 shows the cross-sectional pictures of sprayed TiN coating obtained by optical microscopy. As it is clear in pictures, the thickness of the coating is approximately 40 µm which is much thicker than that of layers deposited by CVD. In reactive plasma spray (RPS) method, thick layers of TiN can be produced with a thickness in the range of a few tens to a few hundreds of micrometres. The thickness of layer increases by increasing the processing time, powder feed rate and decreasing the spray distance [1-2]. The adhesion between coating and substrate was in a good condition as it can be seen in Figure 3.

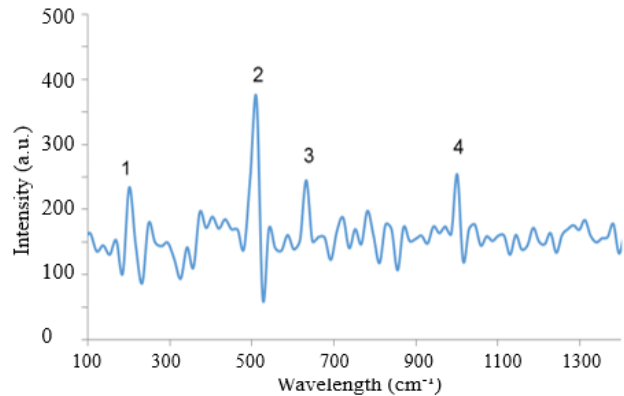

Figure 4: Raman spectroscopy pattern for sprayed coating layer

Raman spectroscopy pattern was recorded to evaluate the chemical composition of deposited TiN layer. The Raman pattern shown in Figure 4 indicates typical titanium nitride films deposited on titanium substrate. Sample Raman scattering spectrum shows some distinguished picks. Pick number 1 and 2 relate to TA and TO mode of titanium nitride Raman spectrum, respectively, and refer to the first order Raman vibration. Another two picks are relating to second order Raman scattering. Pick number 3 relate to 2LA and pick number 4 is associated with 2TO.

The observed picks and unseen picks like LA at first order are resulted from substrate excitation and titanium substrate spectra interference with created titanium nitride film spectra. Shapes and positions of these picks could be influenced by type of lasers and detectors used. As detector resolution increases, picks sharpness degrades. Dispersion intensity at acoustic branches relates to titanium ions vibration and at optical branches relates to nitrogen ions vibration. Intensity of TO mode is more than that of TA mode. These results show that deposited film is purely titanium nitride, demonstrating this method advantages.

In this work spraying has been performed in free atmosphere which can cause oxidation of titanium particles. It would be possible that oxygen molecules diffuse from air to plasma jet and react with Ti particles. To minimize the air diffusion to plasma jet, substrate was placed as close as possible to torch's nozzle (10 cm) and we increased the nitrogen gas flow rate as much as possible. Increasing the inlet gas flow rate increases the gas velocity in plasma jet and subsequently decrease the air diffusion rate into the plasma jet.

## 5. Conclusion

A DC thermal plasma torch was designed and developed for TiN layer deposition on Ti substrate at atmospheric-pressure conditions. Configuration of the torch and nozzle was designed to make a stable nitrogen plasma jet. These experiments demonstrated the feasibility of producing a thick TiN layer with the thickness of 40 μm. The Raman spectroscopy indicated that the deposited layer is mainly consists of TiN. It can be concluded that with this experimental condition all Ti particles are decomposed to Ti atoms and react with nitrogen atoms inside the plasma jet, and then there is no significant air diffusion to the plasma jet as well, resulting in covering all surface of substrate by TiN molecules.

## 6. References

[1] Kobayashi, A. "Formation of TiN coatings by gas tunnel type plasma reactive spraying." Surface and coatings Technology 132 (2) 152-157 (2000).
[2] Bacci, T., L. Bertamini, F. Ferrari, F. P. Galliano, and E. Galvanetto. "Reactive plasma spraying of titanium in nitrogen containing plasma gas." Materials Science and Engineering: A 283 (1) 189-195 (2000).

[3] Shieu, F. S., L. H. Cheng, Y. C. Sung, J. H. Huang, and G. P. Yu. "Microstructure, chemistry and coating properties of ion-plated TiN on type 304 stainless steel." Materials chemistry and physics 50 (3) 248-255 (1997).

[4] Logothetidis, S. B. A. A. U. O. T., I. B. A. A. U. O. T. Alexandrou, and S. B. A. A. U. O. T. Kokkou. "Optimization of TiN thin film growth with in situ monitoring: the effect of bias voltage and nitrogen flow rate." Surface and Coatings Technology 80 (1-2) 66-71 (1996).

[5] Maheo, D., and J. M. Poitevin. "Structure of TiN films deposited on heated and negatively biased silicon substrates." Thin Solid Films 215 (1) 8-13 (1992).

[6] Shahsavan, Martia, and J. Hunter Mack. "Numerical study of a boosted HCCI engine fueled with n-butanol and isobutanol." Energy Conversion and Management 157 (2018): 28-40.

[7] Shahsavan, Martia, and John Hunter Mack. "The effect of heavy working fluids on hydrogen combustion." 10th U. S. National Combustion Meeting, College Park, Maryland, engrxiv, (2017).

[8] Shahsavan, Martia, and John Hunter Mack. "Mixedness Measurement in Gaseous Jet Injection." American Society for Engineering Education Northeast Section, Lowell, Massachusetts, engrxiv (2017).

[9] Shahsavan, Martia, Mohammadrasool Morovatiyan, and John Hunter Mack. "The Influence of Mixedness on Ignition for Hydrogen Direct Injection in a Constant Volume Combustion Chamber." Spring Technical Meeting Eastern States Section of the Combustion Institute, State College, Pennsylvania, engrxiv, (2018).

[10] Morovatiyan, Mohammadrasool, Martia Shahsavan, and John Hunter Mack. "Development of a Constant Volume Combustion Chamber for Material Synthesis." Spring Technical Meeting Eastern States Section of the Combustion Institute, State College, Pennsylvania, engrxiv, (2018).

[11] Udomkan, Nitinai, Vilasinee Sutorn, Pichet Limsuwan, and Pongtip Winotai. "Properties of Titanium Nitride Film Coated on Stainless Steel 304." KASETSART JOURNAL 209 (2003).

[12] Mao, Zhengping, Jing Ma, Jun Wang, and Baode Sun. "Properties of TiN-matrix coating deposited by reactive HVOF spraying." Journal of Coatings Technology and Research 6 (2) 243-250 (2009).

[13] Elahi, Rasool, Mohammad Passandideh-Fard, and Alireza Javanshir. "Simulation of liquid sloshing in 2D containers using the volume of fluid method." Ocean Engineering 96 (2015): 226-244.

[14] Rezaeimoghaddam, Mohammad, Rasool Elahi, MR Modarres Razavi, and Mohammad B. Ayani. "Modeling of Non-Newtonian Fluid Flow Within Simplex Atomizers." In ASME 2010 10th Biennial Conference on Engineering Systems Design and Analysis, pp. 549-556. American Society of Mechanical Engineers, 2010.

[15] Vaz, F., P. Cerqueira, L. Rebouta, S. M. C. Nascimento, E. Alves, Ph Goudeau, J. P. Riviere, K. Pischow, and J. De Rijk. "Structural, optical and mechanical properties of coloured TiN x O y thin films." Thin Solid Films 447, 449-454 (2004).

[16] Javanshir, Alireza, Rasool Elahi, and Mohammad Passandideh Fard. "Numerical Simulation of liquid Sloshing with baffles in the fuel container." In The 12th Iranian Aerospace Society Conference. 2013.

[17] Numerical study of the heat transfer characteristics of a turbulent jet impinging on a cylindrical pedestal

[19] Javanshir, Alireza, Rasool Elahi, and Mohammad Passandideh Fard. "Numerical Simulation of liquid Sloshing with baffles in the fuel container." In The 12th Iranian Aerospace Society Conference. 2013.

[20] Moghiman, Mohammad, Maryam Moeinfar, Rasool Elahi, and Seyed Morteza Abdollahian. "Numerical study of the heat transfer characteristics of a turbulent jet impinging on a cylindrical pedestal." In The Second TSME International Conference on Mechanical Engineering. 2011.

[21] Huang, Haobo, Rachael Howland, Ertan Agar, Mahnaz Nourani, James A. Golen, and Patrick J. Cappillino. "Bioinspired, high-stability, nonaqueous redox flow battery electrolytes." Journal of Materials Chemistry A 5, no. 23 (2017): 11586-11591.

[22] Stoiber, M., S. Perlot, C. Mitterer, M. Beschliesser, C. Lugmair, and R. Kullmer. "PACVD TiN/Ti–B–N multilayers: from micro-to nano-scale." Surface and Coatings technology 177 (2004) 348-354.

[23] Motte, P., M. Proust, J. Torres, Y. Gobil, Y. Morand, J. Palleau, R. Pantel, and M. Juhel. "TiN-CVD process optimization for integration with Cu-CVD." Microelectronic engineering 50 (2000) 369-374.

[24] Yeom, Hwasung, Benjamin Hauch, Guoping Cao, Brenda Garcia-Diaz, Michael Martinez-Rodriguez, Hector Colon-Mercado, Luke Olson, and Kumar Sridharan. "Laser surface annealing and characterization of Ti 2 AlC plasma vapor deposition coating on zirconium-alloy substrate." Thin Solid Films 615 (2016) 202-209.

[25] Mohsenian, Sina, Mahdieh Sadat Esmaili, Jafar Fathi, and Babak Shokri. "Hydrogen and carbon black nano-spheres production via thermal plasma pyrolysis of polymers." International Journal of Hydrogen Energy 41, no. 38 (2016): 16656-16663.

[26] Mohsenian, Sina, Mahdieh S. Esmaili, Babak Shokri, and Mohammad Ghorbanalilu. "Physical characteristics of twin DC thermal plasma torch applied to polymer waste treatment." Journal of Electrostatics 76 (2015): 231-237.

[27] Shaabani, Ahmad, Mohammad Sadegh Laeini, Sina Mohsenian, Babak Shokri, and Ronak Afshari. "Thermal Plasma-processed Natural Hydroxyapatite-MnO2 Nanoparticles as a Reusable and Green Heterogeneous Catalyst for Aerobic Oxidation of Benzylic Alkyl Arenes and Alcohols." Organic Chemistry Research 3, no. 2 (2017): 191-204.

[28] Fathi, J., S. Mohsenian, M. Shafie, H. Mehdikia, and B. Shokri. "Synthesis and analysis of titanium nitride thin film in atmospheric thermal plasma torch." In Plasma Sciences (ICOPS), 2015 IEEE International Conference on, pp. 1-1. IEEE, 2015.

[29] Feng, Wenran, Dianran Yan, Jining He, Guling Zhang, Guangliang Chen, Weichao Gu, and Size Yang. "Microhardness and toughness of the TiN coating prepared by reactive plasma spraying." Applied Surface Science 243 (2005) 204-213.